\documentclass[bibyear,structabstract]{aa} 
\usepackage{natbib}
\bibpunct{(}{)}{;}{a}{}{,} 

\usepackage{graphicx}
\usepackage{txfonts}
\usepackage{subfigure}
\usepackage{hyperref}

\let\bibfont\relax

\begin{document}

\title{High-resolution spectroscopy of flares and CMEs on AD Leo\thanks{Based on observations obtained at the Th\"{u}ringer Landessternwarte Tautenburg, Germany}
}

\authorrunning{Priscilla Muheki et al.}
\titlerunning{Flares and CMEs on AD Leo}
\author{
P. Muheki \inst{1,\,2}\and
E.W. Guenther\inst{1}\and
T. Mutabazi \inst{2}
\and
E. Jurua \inst{2}
          }
   \institute{
   Th\"uringer Landessternwarte Tautenburg, Sternwarte 5, 07778
Tautenburg, Germany 
         \and
         Mbarara University of Science and Technology,
P.O Box 1410, Mbarara, Uganda
\\\\
              \email{pmuheki@must.ac.ug; guenther@tls-tautenburg.de; ejurua@must.ac.ug}\\
}
  \abstract 
{Flares and coronal mass ejections (CMEs) are important for
  the evolution of the atmospheres of planets and their potential
  habitability, particularly for planets orbiting M stars at a distance $\rm < 0.4\, AU$. Detections of CMEs on these stars have been sparse, and previous studies have therefore modelled their occurrence frequency by scaling up solar relations.  However, because the topology and strength of the
  magnetic fields on M stars is different from that of the Sun, it is not
  obvious that this approach works well. }
{ We used a large number of high-resolution spectra to study flares, CMEs, and their dynamics  of the active M dwarf star \object{AD\,Leo}. The results can
  then be used as reference for other M dwarfs.}
{We obtained more than 2000 high-resolution spectra ($R\sim 35000$) of the highly
  active M dwarf \object{AD\,Leo,} which is viewed nearly pole on. Using these data, we studied the behaviour of the spectral lines $\rm H_{\alpha}$, $\rm H_{\beta}$, and $\rm
  He\,I\,5876$ in detail and investigated asymmetric features that might be Doppler signatures of CMEs. }
{We detected numerous flares. The largest flare emitted $8.32 \times 10^{31}$
  erg in $\rm H_{\beta}$ and $2.12 \times 10^{32}$ erg in $\rm H_{\alpha}$. Although
  the spectral lines in this and other events showed a significant
  blue asymmetry, the velocities associated with it are far below the
  escape velocity. }
{Although AD Leo shows a high level of flare
  activity, the number of CMEs is relatively low. It is thus not
  appropriate to use the same flare-to-CME relation for M dwarfs as
  for the Sun. }
\keywords{Techniques: spectroscopic -- 
          Stars: activity -- 
          Stars: flares -- 
          Stars: late--type --  
          Stars: individual: AD\,Leo 
          }

\maketitle
%

\section{Introduction}
\label{sectI}
Observations with the Kepler and CoRoT satellites have shown that
extrasolar planets are much more diverse than the planets in our
Solar System \citep{lammer11}.  This large diversity especially in density is particularly striking for
planets in the mass range between 1 and 15 $\mbox{M}_{\rm{Earth}}$ \citep{Hatzes15}. In this mass range, planets can have low densities, for example, \object{Kepler-51} \citep[0.05 $\rm g\,cm^{-3}$,~][]{masuda14}, or very high densities (\object{K2-106,} \citealp[$13.1^{-3.6}_{+5.4}~\rm g\,cm^{-3}$,~][]{guenther17}; \object{K2-229b,} \citealp[$8.9 \pm 2.1~ \rm g\,cm^{-3}$,][]{santerne18}; \object{Kepler-107c,} \citealp[$12.65 \pm 2.45~ \rm g\,cm^{-3}$,~][]{bonomo19}). 
This large diversity in densities is obviously related to different compositions of the planets.  The high-density planets must be rocky and
cannot have a dense, extended atmosphere. In contrast to this,
planets with low density must correspondingly have extended
atmospheres that mostly consist of $\rm H_2/He$ \citep{lissauer11, lammer11}.  The large
variation in the structure of planets must be related to their formation
and evolution. High-density planets either never had an extended $\rm
H_2/He$ atmosphere or lost it when they were young \citep{lissauer11}. Perhaps both scenarios play
a role.

Many studies \citep[e.g.~][]{lammer8,lammer14} have shown that atmospheric loss processes are important
for the evolution of planets. Low-mass planets such as \object{CoRoT-7b,}
which orbits within 0.02 AU, could lose their atmosphere by extreme ultra-violet (EUV) and X-ray driven
photoevaporation or in a boil-off event \citep[e.g.~][]{lammer13, kislyakova13, erkaev16, roettenbacher17}.  For
example, \citet{roettenbacher17} showed for the
\object{TRAPPIST-1} system that the environment has potentially led to the
erosion of the atmospheres of planets in the habitable zone through
EUV stellar emission.  \citet{bolmont17} further showed that TRAPPIST-1b and c may
have lost about 15 Earth oceans, but \object{TRAPPIST-1d} about less than one Earth ocean. These
planets could thus  only have water if the original water contents were
one or a few Earth oceans higher than that. However, the simulation by
\citet{tian15} showed that M stars are expected to have
two types of planets in the habitable zones: ocean planets
without continents, and desert planets, on which there is orders of
magnitude less surface water than on Earth. Most of the observational
studies \citep[e.g.~][]{kevin12,roettenbacher17, bolmont17} have concentrated on X-ray and EUV radiation (XUV radiation) of
the stars in quiescence, but flares and coronal mass ejections (CMEs)
certainly play a role as well \citep{lammer7, lammer09}.

Flares affect the potential habitability of planets in many ways, and these effects are particularly important for planets orbiting M stars. According to \citet{tuomi19}, three planets on average orbit M stars. These stars have a relatively high activity, but the potentially habitable planets orbit at considerably small distance ($\rm \sim 0.03 AU - 0.4 AU$)
\citep[e.g.~][]{segura10, vida17, howard18, tuomi19}. The importance of
flares on M dwarfs is underlined by the fact that they might dominate the far-UV (FUV) energy budget \citep{loyd18}. An interesting example for the effect of flares on the habitability of planets is TRAPPIST-1 \citep{gillon16}. \citet{vida17} showed for this star that the flare activity is so high that the atmospheres of the planets orbiting in the habitable zone may be continuously altered, making them less favourable for hosting life. Another interesting example is Proxima Centauri. \citet{kielkopf19} observed a superflare on this star in which the energy of the flare was 100 times the star's luminosity, and such large flares may endanger life on a planet in the habitable zone.

On the other hand, the erosion of planetary atmospheres caused by CMEs (due to ion-pick
up) depends on the plasma densities and flow velocities of the
CMEs, and the magnetic moments of the planets \citep{lammer9}. 
As shown by \citet{kay16}, exoplanets around M dwarfs would require magnetic fields of between tens and
hundreds of Gauss to protect them against CMEs, but because close-in planets are tidally locked \citep{lammer7, barnes17}, it is
generally thought that their magnetic moments are relatively
low. According to \citet{lammer14}, planets with an
Earth-like structure orbiting an M star may lose their entire
atmosphere as a result of CMEs. This is based on the assumption that the CME rate is 100 times higher than that of the Sun because the X-ray brightness of such stars is 100 times higher. However, it is not certainly known what the CME rate of M dwarfs really is. 

Previous studies \citep[e.g.][]{aarnio12, drake13, leitzinger14, osten15, odert17, crosley18} used the correlations between solar flare energies and CME parameters to estimate possible occurrence frequencies of CMEs in M stars. However, it is still uncertain whether it is correct to use the solar analogy for these very active stars.
Solar CMEs originate from regions that have a very
specific magnetic field geometry, that is, regions of complex magnetic topology and high field strength \citep{aarnio12}. The topology of fully convective M dwarfs is very complex and divergent \citep{morin08, morin10}. 
Rapidly rotating M dwarfs have several kiloGauss large-scale magnetic fields at high latitudes, and their coronae are dominated by star-size large hot loops  \citep{cohen17}. In addition to this, active M dwarfs with dipole-dominated axisymmetric field topologies can undergo a long-term global magnetic field variation \citep{lavail18}. 
Because the topology of the magnetic fields in M dwarfs may differ from that of the Sun, and because the release of matter in CMEs requires certain field geometries of the overlying magnetic field \citep{odert17}, it is not obvious that it is possible to use the
solar analogy to estimate the effect of CMEs on planets orbiting M
dwarfs. 
A much better approach is to study energies, velocities, and
masses of CMEs during flare events on active M dwarfs and to extrapolate these results to the less active M dwarfs.
\citet{loyd18} showed that when flare energies were normalised by the star's quiescent flux, the active and inactive samples exhibited identical slopes of the cumulative flare frequency distribution. Thus it is possible to study active M dwarfs and extrapolate the results to less active ones.

The spectroscopic signature of CMEs are blue-shifted components where
the velocity shift exceeds the escape velocity of the star \citep[][and references there in]{vida19}.
CMEs have been observed on M dwarfs \citep[e.g.~][]{houdebine90, cully94, gunn94, guenther97, vida16}, but all these were serendipitous events and thus are not enough for a statistical inference on their occurrence rate.
\citet{vida16} showed that the observed CME rate is much lower than the expected rate derived from
the CME--flare ratio on the Sun. 
Given the expected low rate of such events, we need to observe the stars for a long time, but even then it is useful to observe a particularly active star in order to observe more flares and CMEs.   

For this study we selected the highly active M4 dwarf
\object{AD\,Leo} (=GJ 388) \citep{keenan89, lurie14}.
A spectropolarimetric study by \citet{morin08} showed
that the star is seen nearly pole-on. The field is 
poloidal with or without the presence  of a toroidal component that
accounts for 1-5\%\ of the magnetic energy. This star's high activity and the field geometry orientation increase the chances for observing CMEs compared to other M dwarfs. 

\object{AD Leo}  has been the subject of a number of flare
studies. On April 12, 1985, \citet{hawley91} observed a flare that released 
$\rm 9.8\times 10^{-8}\,erg\, cm^{-2}$ in $\rm H_\alpha$
and $\rm 1245\times 10^{-8}\,erg\, cm^{-2}$ in total. \citet{hawley03} studied flares on AD Leo using multi-wavelength simultaneous photometry and high-resolution spectroscopy. They studied 8 sizeable flares in four nights of observation, giving insights into the correlation between continuum and line emission in the optical and ultraviolet wavelengths.
A detailed spectroscopic study of the flares
on \object{AD Leo} was also presented by
\citet{crespo06}.
During the 16.1 hours of monitoring time,
the authors obtained 600 spectra and observed 
14 flares. The energies of the flares
range from 1.5 to $8.2 \times 10^{29}$ erg in $\rm H_\beta$.
The resolution of the blue spectra  that include 
$\rm H_\beta$ was $\rm R\sim4400,$ and that of the red spectra 
that cover $\rm H_\alpha$ was $\rm R\sim5600$. While these
spectra allow measuring the line fluxes, the resolution
is not high enough to study the subtle changes of the line 
profiles. \citet{buccino14} studied the chromospheric activity of AD Leo using medium-resolution echelle spectra taken over a period of 13 years. However, they focused on studying the variations in the activity, thereby neglecting spectra that had flares.

Extreme ultra-violet radiation causes the atmospheric loss processes by heating the upper atmospheres of the planets. However, it is difficult to observe and monitor active M stars for a long time at these wavelengths. Moreover, stellar CMEs can be detected using spectroscopy by studying Doppler shifts and line asymmetries with enhanced blue wings of the stellar emission lines. When the component  is blue-shifted by more than the escape velocity, we can be certain that this material is ejected by the star. We  therefore used optical observations to study the flare rates, their energies, and the CMEs, which can then be used in models to calculate the atmospheric loss rates. The Balmer lines are ideally suited for this purpose because we know from solar observations that these lines are closely related to the XUV radiation because the $\rm H_{\alpha}$ loops are located just below the X-ray loops and thus are heated by XUV radiation from above \citep{butler93}.\\We present here a comprehensive study of the flare activity of \object{AD\,Leo} using high-resolution spectra taken over a long observing period. This enabled us to detect much smaller wavelength shifts and thus gain more insight into CME dynamics on this star.

\section{Observations and data reduction}
\label{sectII}

\object{AD\,Leo} is one of the stars that was monitored during the flare-search
program of the Th\"uringer Landessternwarte. For this program we used the 2m Alfred Jensch telescope of the Th\"uringer
Landessternwarte Tautenburg, which is equipped with an \'echelle
spectrograph with a resolving power of $\rm R \sim 35000,$  and we used the $2^{\prime\prime }$ slit. The spectra cover the wavelength region from 4536 to
7592 \AA. With a resolution of $8.5 \, \rm km\,s^{-1}$, it is possible to  detect CMEs and also study more subtle changes of the line profiles during such events. Spectra were obtained in various campaigns from November
2016 until February 2019. Observations were typically scheduled for one
week per month, when the object was visible. When possible,
\object{AD\,Leo} was observed continuously throughout the night. The
exposure times were initially set to 600\,s, but after the large event
observed in February 2018, we decided to reduce the exposure time to
300\,s in order to obtain a better sampling of the rise times of the flares. We thus obtained 862 spectra with 600\,s exposure time and a monitoring time of 143.7\,h. Since April 2018, we have monitored \object{AD\,Leo} for 95\,h, obtaining 1140 spectra  with 300\,s exposure time. The total monitoring time was thus 238.7\,h in 57 nights of observation. A log of the observations is given in Table \ref{tab1}. 
\begin{figure}
 \includegraphics[width=0.5\textwidth]{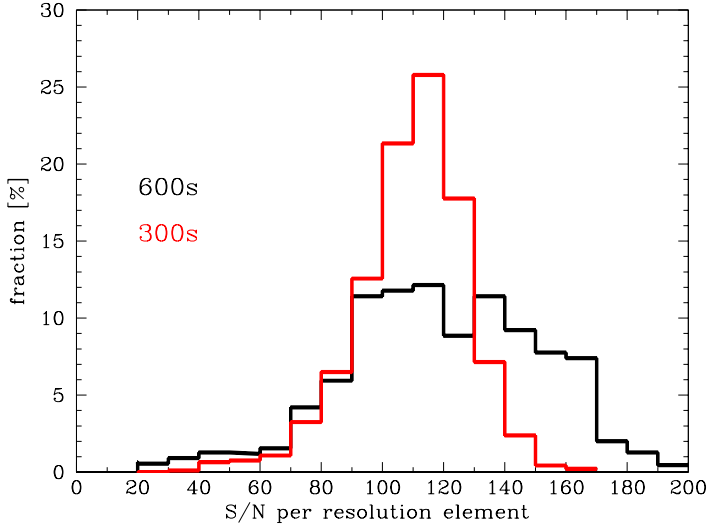}
\caption{\label{snr}Variation of the S/N per resolution element of the spectra.}
\end{figure}
In Fig. \ref{snr} we show the variation of the signal-to-noise ratio (S/N) per resolution element. 

The spectra obtained were bias-subtracted, flat-fielded, the scatter light
removed and extracted using standard IRAF routines. Spectra taken with
a ThAr lamp were used for the wavelength calibration.
\section{Results}
\label{sectIII}
\subsection{Flare observations}
\object{AD\,Leo} shows a very high level of chromospheric activity, especially in the Balmer lines ($\rm H_{\alpha}$ and $\rm H_{\beta}$) and the $\rm He\, I\, D_3$ , and we therefore studied the behaviour of these three lines. In order to detect flares in \object{AD\,Leo}, we measured the fluxes in the $\rm H_{\alpha}$, $\rm H_{\beta}$ , and $\rm He\,I\, D_3$ lines and used them to construct light curves. 
\begin{figure}
 \includegraphics[width=.5\textwidth]{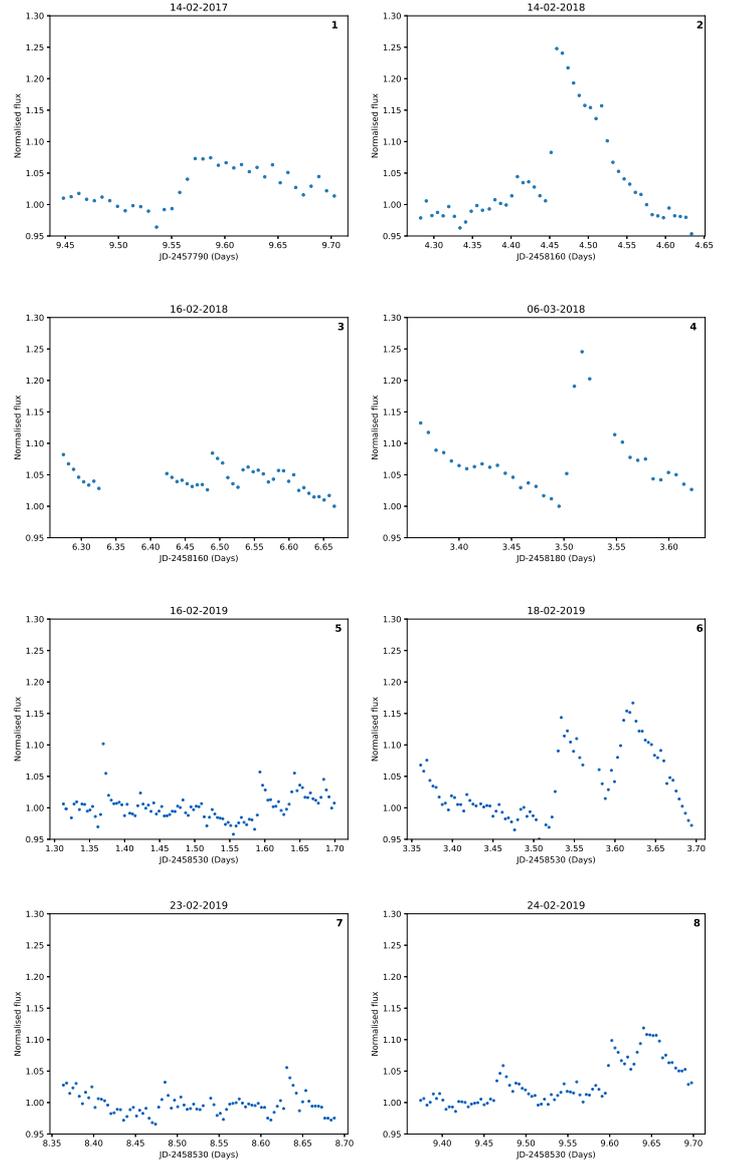}
\caption{\label{lightcurves}Temporal evolution of the flux of the $\rm H_{\alpha}$ line, showing light curves of the strongest observed flares.}
\end{figure}

\begin{figure}
 \includegraphics[width=.4\textwidth]{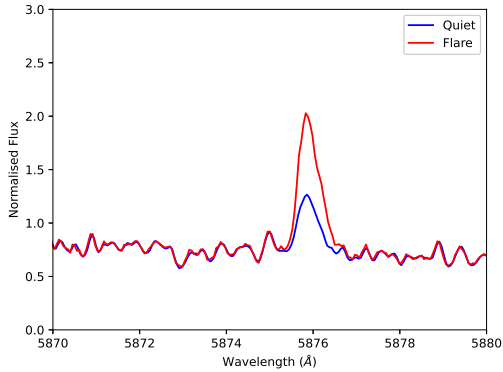}
\caption{\label{helium}Extracted spectrum of AD\,Leo showing the enhancement in the He\,I\,5876 emission line during the flare in panel 2 of Fig.\,\ref{lightcurves}.}
\end{figure}

Examples of the relative flux variation in $\rm H_{\alpha}$ of the flares are shown in Fig. \ref{lightcurves}. As is typical for flares, not all the events show the classical light curve (panel 2) of  a sudden increase and slow decrease, but often a more complex structure (panel 1). Figure \,\ref{lightcurves} shows nights in which multiple events occured (e.g. panels 5-8).\\ The enhancement in the He\,I\,5876 emission line (see Fig.\,\ref{helium}) shows that a plasma that is much hotter than the photosphere is present during these events. The He\,I\,5876 emission line occurs at about 20000 K \citep{giampapa78}. However, we did not detect the He\,II\,4686 line, which requires a temperature of 30000 K to be excited \citep{lamzin89}. This means that the plasma was not hot enough to excite this line.
In addition to the unfavourable weather conditions that interrupted the observations especially in 2016, the activity seems to be lower in 2016 and 2017 than in 2018 and 2019. There is an increase in the activity levels beginning in February 2018 (see Table \ref{tab1}). 

We obtained an average flare rate of $\approx 0.092$ flares per hour, ignoring nights with fewer than six spectra. However, as we explain in section \ref{energies}, flares have a power-law distribution, which means that the number of flares that are detected depends on the sensitivity of the measurements. The flare rates of solar observations are much higher than those of stellar observations because the Sun is much brighter and we cannot spatially resolve the stellar disc for stellar observations.
When we consider only 2018 and 2019, we obtain a slightly higher rate of $\approx 0.107$ flares per hour. 
Generally, the activity levels are much lower than those obtained in other
studies e.g., f = 0.57 \citep{pettersen84}, $\rm f\,> 0.71$ \citep{crespo06} and $\rm f\sim 0.1$ \citep{hunt12}, where f is the frequency of flares per hour.

\citet{buccino14} obtained two possible activity cycles for \object{AD\,Leo}: one of seven years, and another less significant cycle of two years. The seven-year cycle reached a minimum in 2007, and the authors predicted that the next minimum would be in 2015 and that the stellar activity would be even lower. This might explain why the activity levels were low in 2016 to 2017 and have started to increase beginning in 2018. 

\subsection{Flare energies}
\label{energies}
In order to estimate the energy released during a flare, we integrated the luminosity over the flare duration. The absolute line fluxes were computed from the flux of each line and its nearby continuum. The continuum flux near the emission lines  was determined using direct scaling of the flux-calibrated spectrum of \object{AD\,Leo} from \citet{cincunegui04}, assuming the same brightness of the star over time because \object{AD\,Leo} has very small photometric variations of $\approx$ 0.04 mag \citep{buccino14}. If the continuum were to increase during the flare, the relative strength of the line would decrease. By assuming the continuum is constant, we therefore obtain a lower limit for the line flux and thus the flares might be more energetic than estimated. To obtain the luminosities, we used the distance to the star of $4.9660\pm 0.0017 $\,pc from \textit{Gaia} \citep{gaia18}. The estimated values of the energy released in each observed flare are also shown in Table \ref{tab1}. The energies given in Table \ref{tab1} are for $\rm H_{\alpha}$, $\rm H_{\beta}$ , and He\,I\,5876.
 The smallest flare we detected had an energy of  $\rm 3.19\pm 0.01 \times 10^{31}\, erg$ in $\rm H_{\alpha}$, $\rm 8.40\pm 0.01 \times 10^{30}$ erg in $\mbox{H}_{\beta}$ and $9.93 \pm 0.05 \times 10^{29}$ erg in He\,I\,5876.
 \begin{table*}
\caption{\label{tab1}Observation log showing nights of observation of \object {AD\,Leo}. The number of hours per night and the corresponding number of spectra obtained are shown in Cols. 3 and 5, respectively. Columns 6, 7, and 8 correspond to the estimated energies for the flares observed in $\rm H_{\alpha}$, $\rm H_{\beta}$ , and He\,I\,5876.}

\begin{tabular}{p{1.7cm} p{3.0cm} p{1.5cm} p{1.5cm} p{1.5cm} p{0.05cm}cccccc}
\hline
\hline
Obs. date & Start--End time (UTC) & Obs  (h) & Exposure time (s) & Number of Spectra &&& \multicolumn{3}{c}{Flare Energy ($10^{31}$erg)}\\

&&&&&&H$_\alpha$&H$_\beta$&He\,I\,5876\\[0.5ex]
\hline
2016-11-19 &01:29 - 03:47 &2.33&600&14&&-&-&-\\
2016-11-20 &00:01 - 00:32 & 0.5&600&3&&-&-&-\\
2016-12-09 &01:17 - 01:47 & 0.5&600&3&&-&-&-\\
2016-12-16 &23:25 - 01:30 & 2&600&12&&-&-&-\\
2016-12-21 &01:23 - 04:55 & 3.5&600&21&&-&-&-\\
2017-02-03&23:54 - 00:15&0.33&600&2&&-&-&-\\
2017-02-04 &22:19 - 00:03 &1.83 &600 &11&&-&-&-\\
2017-02-13 &01:11 - 04:21 &3.17&600&19&&-&-&-\\
2017-02-14{$^\dagger$} &22:37 - 04:44 &6&600&36&&12.05 $\pm$ 0.03& 9.495$\pm$0.077&0.553$\pm$0.004\\
2017-02-15 &22:58 - 04:26 &5.33&600&32&&-&-&-\\
2017-03-10 &20:41 - 21:01&0.33&600&2&&-&-&-\\
2017-03-10 &18:51 - 01:51 &6.83&600&41&&-&-&-\\
2017-03-11 &18:14 - 03:15 &8.67&600&52&&-&-&-\\
2017-03-12 &18:25 - 02:24 &3.83&600&23&&-&-&-\\
2017-04-15 &21:41 - 23:15 &1.67&600&10&&-&-&-\\
2017-05-03&20:55 - 21:25&0.5&600&3&&-&-&-\\
2017-05-06 &19:33 - 22:21 &2.83 &600&17&&-&-&-\\
2017-05-07 &19:17 - 20:20 &1.17 &600&7&&-&-&-\\
2017-05-09&20:55 - 21:26&0.5&600&3&&-&-&-\\
2017-05-10 &20:16 - 21:19 &1.17 &600&7&&-&-&-\\
2017-05-15$^\dagger$ &20:31 - 21:47 &1.25 &600&8&&1.593$\pm$0.005 & -&-\\
2017-12-24 &02:25 - 03:07 &0.5 &600&3&&-&-&-\\
2017-12-25 &23:53 - 05:07 &3.67 &600&22&&-&-&-\\
2017-12-26 &00:37 - 01:40 &1.33 &600&7&&-&-&-\\
2017-12-28 &22:53 - 00:06&1.25 &600&8&&-&-&-\\
2018-01-05 &01:28 -03:45&2.33&600&14&&-&-&-\\
2018-02-12 &19:25 - 23:23 &2.67 &600&16&&-&-&-\\
2018-02-13$^\dagger$ &18:29 - 04:48 &10 &600&60&&3.155$\pm$0.033&1.154$\pm$0.012&0.154$\pm$0.002\\
2018-02-14 &18:39 - 03:04 &8.17 &600&49&&5.854$\pm$0.031 & 2.195$\pm$0.036&0.443$\pm$0.007\\
&&&&&&21.2$\pm$0.4 & 8.323$\pm$0.319&1.998$\pm$0.077\\
2018-02-16$^\dagger$ &18:26 - 03:50 &7&600&42&&19.472$\pm$0.079&5.11$\pm$0.06&0.428$\pm$0.002\\
2018-02-18 &18:25 - 01:03 &5.83 &600&35&&11.909$\pm$0.082&3.715$\pm$0.069&0.391$\pm$0.003\\
2018-02-26 &21:04 - 04:17 &7 &600&42&&14.599$\pm$0.061&3.081$\pm$0.021&0.499$\pm$0.004\\
2018-02-27 &18:22 - 04:05 &9.33 &600&56&&-&-&-\\
2018-02-28 &18:19 - 04:20 &9.67 &600&58&&-&-&-\\
2018-03-01 &18:14 - 04:00 &9.33 &600&56&&-&-&-\\
2018-03-04 &00:15 - 02:21 &2.17 &600&13&&-&-&-\\
2018-03-05 &20:35 - 02:47 &5.67 &600&34&&8.406$\pm$0.135&5.031$\pm$0.255&0.536$\pm$0.013\\
2018-03-08 &23:03 - 03:46 &3.5 &600&21&&-&-&-\\
2018-03-24 &18:35 - 02:16 &6.92 &300&83&&-&-&-\\
2018-03-29 &19:47 - 01:36 &4.58 &300&55&&-&-&-\\
2018-04-23 &19:34 - 23:21 &3.5 &300&42&&-&-&-\\
2018-04-27$^\dagger$ &19:43 - 22:22 &2.5 &300&30&&1.913$\pm$0.022&0.691$\pm$0.035&0.069$\pm$0.004\\
2018-04-29 &19:42 - 22:21 &2.5 &300&30&&-&-&-\\
2018-05-01 &19:58 - 22:37 &2.5&300&30&&-&-&-\\
2018-05-04 &19:56 - 22:35 &2.5 &300&30&&-&-&-\\
2018-05-05 &19:59 - 22:11 &2.08 &300&25&&-&-&-\\
2018-05-07 &19:52 - 22:31 &2.5 &300&30&&-&-&-\\
2019-02-13 &23:28 - 04:31 &4.83&300&58&&-&-&-\\
2019-02-14 &19:12 - 04:38 &8.67&300&104&&8.518$\pm$0.025&2.234$\pm$0.013&0.524$\pm$0.002\\
2019-02-15 &19:31 - 04:53 &8.67&300&104&&-&-&-\\
2019-02-16&19:22 - 04:38 &8.58&300&103&&4.505$\pm$0.041&1.12$\pm$0.01&0.123$\pm$0.001\\
&&&&&&3.19$\pm$0.01&0.84$\pm$0.01&0.099$\pm$0.0005\\
&&&&&&3.860$\pm$0.023&8.61$\pm$0.01&0.126$\pm$0.001\\
&&&&&&5.531$\pm$0.019&1.364$\pm$0.012&0.192$\pm$0.0008\\
2019-02-17$^\dagger$&19:23 - 04:18 &8.33&300&100&&6.628$\pm$0.039&1.97$\pm$0.02&0.300$\pm$0.003\\
&&&&&&2.884$\pm$0.019&0.669$\pm$0.008&0.100$\pm$0.0003\\
2019-02-18&20:31 - 04:31&7&300&84&&15.479$\pm$0.012&3.045$\pm$0.049&0.408$\pm$0.002\\
2019-02-23&20:35 - 04:22&7.08&300&85&&5.939$\pm$0.017&1.385$\pm$0.008&0.466$\pm$0.002\\
&&&&&&5.295$\pm$0.032&1.235$\pm$0.018&0.419$\pm$0.005\\
2019-02-24$^\dagger$&20:5  04:36&7.17&300&86&&5.488$\pm,$0.023&1.582$\pm$0.013&0.145$\pm$0.0009\\
&&&&&&11.083$\pm$0.048&2.636$\pm$0.043&0.292$\pm$0.002\\
2019-02-25&20:39 - 02:12&5.08&300&61&&-&-&-\\
\hline
\end{tabular}
~\\
$\dagger$ indicates nights in which the value of the energy is a lower limit as only the impulsive phase or part of the  gradual phase was observed. 

\end{table*}
 
\begin{figure}

 \includegraphics[width=.5\textwidth]{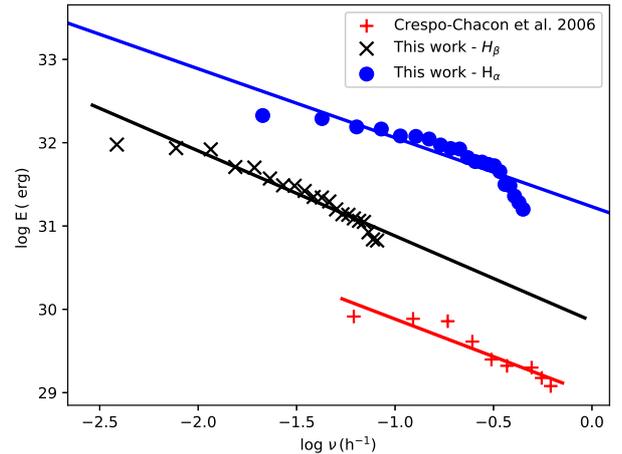}
\caption{\label{cum}Cumulative flare frequency distribution for \object{AD\,Leo}. The solid lines represent our best linear least-squares fit to each of the Flare Frequency Distributions. Our observations of flares in $\rm H_{\beta}$ (black crosses) show a similar value of $\beta$ ($-0.97\pm0.06$) to that of \citet{crespo06} i.e., $\beta = -0.96\pm0.14$ (red plus symbols) who also observed in $\rm H_{\beta}$.}
\end{figure}

 Fig.\ref{cum} shows the cumulative frequency distribution diagram for the observed flares.
The cumulative frequency distribution of flares follows a Pareto distribution and can be fitted by a linear relation,
\begin{equation}
 \log\,\nu = a + \beta \log\,E
 \label{eqn1}
,\end{equation}
 where $\nu$ is the cumulative frequency of flares with an energy E or greater and $\beta$ is the power-law exponent. The more energetic flares occur less frequently, as shown by several photometric studies \citep[e.g.][]{lacy76, pettersen84, hunt12}. The lines in Fig.\,\ref{cum} are the formal fits of a straight line (Eqn.\,\ref{eqn1}) to the data to give a scale.
  
 We obtained a value of $\beta = -0.98\pm 0.21$ for the flares in  $\rm H_{\alpha}$ and $\beta = -0.97\pm0.06$ in $\rm H_{\beta}$. This is close to what has previously been obtained for \object{AD\,Leo}, for instance, $-0.82\pm0.27$ \citep{lacy76}, 0.85 \citep{audard00}, and   $-0.68\pm0.16$ \citep{hunt12}. However, our fit is steeper probably because we sample low-energy events better. Figure\,\ref{cum} shows that the last five points bend over. The cumulative frequency distribution bends over probably because we miss some flares close to the sensitivity limit. When we ignore these points in the fitting, the slope changes by $\sim 0.58$ for the flares in $\rm H_{\alpha}$.
 
 In Fig.\,\ref{cum} we also show the flare frequency distributions for the flares in $\rm H_{\beta}$ for this work and that of \citet{crespo06}. \citet{crespo06} observed \object{AD\,Leo} for 16.2 h in $\rm H_{\beta}$ , and despite the longer period of observation for this work, the two data sets show substantial agreement in the distribution of flares on \object{AD\,Leo}. We obtained a value of $\beta = -0.96\pm0.14$ for the data obtained from \citet{crespo06}, which is similar to the value we obtained for the flares in $\rm H_{\beta}$ for this work. \object{AD\,Leo} was on average more active during our observation time than on April 2-5, 2001 when \citet{crespo06} observed it. Because the last maximum as observed by \citet{buccino14} was in 2011, it is expected that there also was a maximum in about 2018, which falls in our time of observations.
 
 According to \citet{gizis17}, the slope of the fit is greatly influenced by the energy range of the flares. A maximum likelihood estimation gives a better prediction of the value of $\alpha,$ especially for heavy tailed power-law distributions. The value of $\alpha$ is obtained from the equation
 \begin{equation}
 \alpha = 1+ n\big[\sum_{i=1}^n \ln\frac{\mbox{E}_i}{\mbox{E}_{min}}\big]^{-1}
 \label{eqn2}
.\end{equation}
From \,$\beta =1-\alpha$ \citep{hawley14}, we obtained values of $\beta$ = $ -1.45\pm 0.26$ for the flares in $\rm H_{\alpha}$, $-1.12\pm 0.35$ in $\rm H_{\beta}$ , and $-1.01\pm 0.44$ for data from \citet{crespo06}. The discrepancy in the values to those obtained from the least-squares fitting especially for $\rm H_{\alpha}$ is probably due to the fact that it has a heavy tailed distribution. This may be attributed not only to the closeness to the sensitivity limit, but also to the fact that some values of the energy were a lower limit when only the impulsive phase was observed. When we account for the bias caused by the smallness of the samples, as in \citet{gizis17}, we obtain new values of $\beta = -1.23\pm0.33,~ -0.92\pm0.26,~ \rm and ~ -0.57\pm0.33,$ respectively. Within the error ranges, these values are rather similar to those obtained from the least-squares fitting. Table\,\ref{tab3} gives a summary of the $\beta$ values obtained using the different techniques.
\begin{table}
\caption{Summary of the $\beta$ values obtained using the different techniques. l.s.f is least-squares fitting, and m.l.e is maximum likelihood estimation. $\rm ^a$ implies least-squares fitting considering all data points, $\rm ^b$ is least-squares fitting ignoring 5 points with lowest energy, $\rm ^c$ implies maximum likelihood estimation with bias and $\rm ^d$ is maximum likelihood estimate without bias. $\rm H_{\beta}^+$   corresponds to data from \citet{crespo06} }
  \begin{tabular} {p{0.5cm} p{0.5cm} p{2.0cm} p{2.0cm} p{2.cm} p{1.cm}cccc}
\hline
Method &&\multicolumn{3}{c}{$\beta$ value}\\
&& $\rm H_{\alpha}$ & $\rm H_{\beta}$ & $\rm H_{\beta}^+$\\[0.5ex]
\hline
$\rm l.s.f^a$&&$-0.98\pm0.21$&$-0.97\pm0.06$&$-0.96\pm0.14$\\
$\rm l.s.f^b$&&$-1.56\pm0.13$&$-1.10\pm0.08$&--\\
$\rm m.l.e^c$&&$-1.45\pm0.26$&$-1.12\pm0.35$&$-1.01\pm0.44$\\
$\rm m.l.e^d$&&$-1.23\pm0.33$&$-0.92\pm0.26$&$-0.57\pm0.33$\\

\hline
\end{tabular}
 
 \label{tab3}
 \end{table}

The ratio of the energy in $\rm H_{\alpha}$ to the energy in $\rm H_{\beta}$ does not depend on the energy of the events.
We obtained an average flux ratio of $3.36\pm 0.26$. With this ratio, we are able to correlate the energy in the Balmer lines to that in the X-ray also using the ratio of the energy in $\rm H_{\beta}$ to that in $\rm H_{\gamma}$  \citep[$1.82\pm0.18$,~][]{crespo06}. The luminosity in X-ray can be related to the luminosity in $\rm H_{\gamma}$ from the relation $\rm L_x(0.04~ to~ 2.0~keV)=31.6L_{\rm H_{\gamma}}$ \citep{butler93}.   
We also calculated the average flux ratios of the He\,I\,5876/$\rm H_{\alpha}$ as 0.0445$\pm$0.022 and  He\,I\,5876/$\rm H_{\beta}$ as 0.145$\pm$0.085. There is no correlation observed between the total line flux and these ratios. 
The largest flare was observed in the night of February 14,
2018.  In that night, we monitored \object{AD\,Leo} continuously from
18:39 UT to 03:14 UT. During that time, we obtained 49 spectra with an exposure time of 600 s. 
\begin{figure}
\includegraphics[width = .5\textwidth]{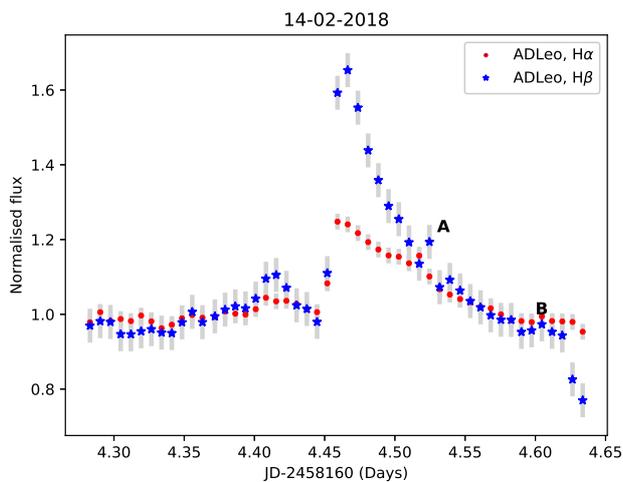}

\caption{Light curves of the flare observed on  February 14, 2018, in $\rm H_{\alpha}$ (filled red circles) and  $\rm H_{\beta}$ (filled blue stars).} 
\label{Fig1}

\end{figure}

The total emission during the flare was
$2.9\times 10^{31}$ erg in $\rm H_{\beta}$ and $2.12\times 10^{32}$ erg in $\rm
H\alpha$. 
As we discuss below, flares of this size certainly affect planetary atmospheres.
During the decay phase of the flare, two events that we may classify as weak events were observed. 
 The first, marked as A in Fig.\,\ref{Fig1}, lasted about $7.2\times10^{3}$ s, and event B lasted about $3\times 10^3$ s. 
These weak post-flare events tend to occur during large CMEs on the Sun as a result of the movement of magnetic loop footpoints that form current sheets and in this way trigger other smaller events \citep{Kovari07}. Pre-flare dips are also observed before the impulsive phase of almost all the flares that were observed, including this large flare. The dips last for about 10--30 min. The duration of these dips is consistent with those presented in literature \citep[e.g.][]{ventura95, leitzinger14}. These dips might be a result of variations in the $\rm H_{\alpha}$ flux due to pre-flare stellar activity \citep{leitzinger14}. CMEs on the Sun are commonly associated with very energetic flares \citep{yashiro09}.

\subsection{Line asymmetries}
In order to detect CMEs, we searched for line asymmetries by visual inspection of the average subtracted spectra. We also measured the position of the line centre at each intensity, thus obtaining accurate radial velocities at each intensity level. We observed line asymmetries (in red and in blue) and broadening of wings in several spectra during flares. Table \ref{tab2} gives a summary of the asymmetries observed and the velocities determined. 

\begin{figure}
\subfigure
{       
\includegraphics[scale=0.5]{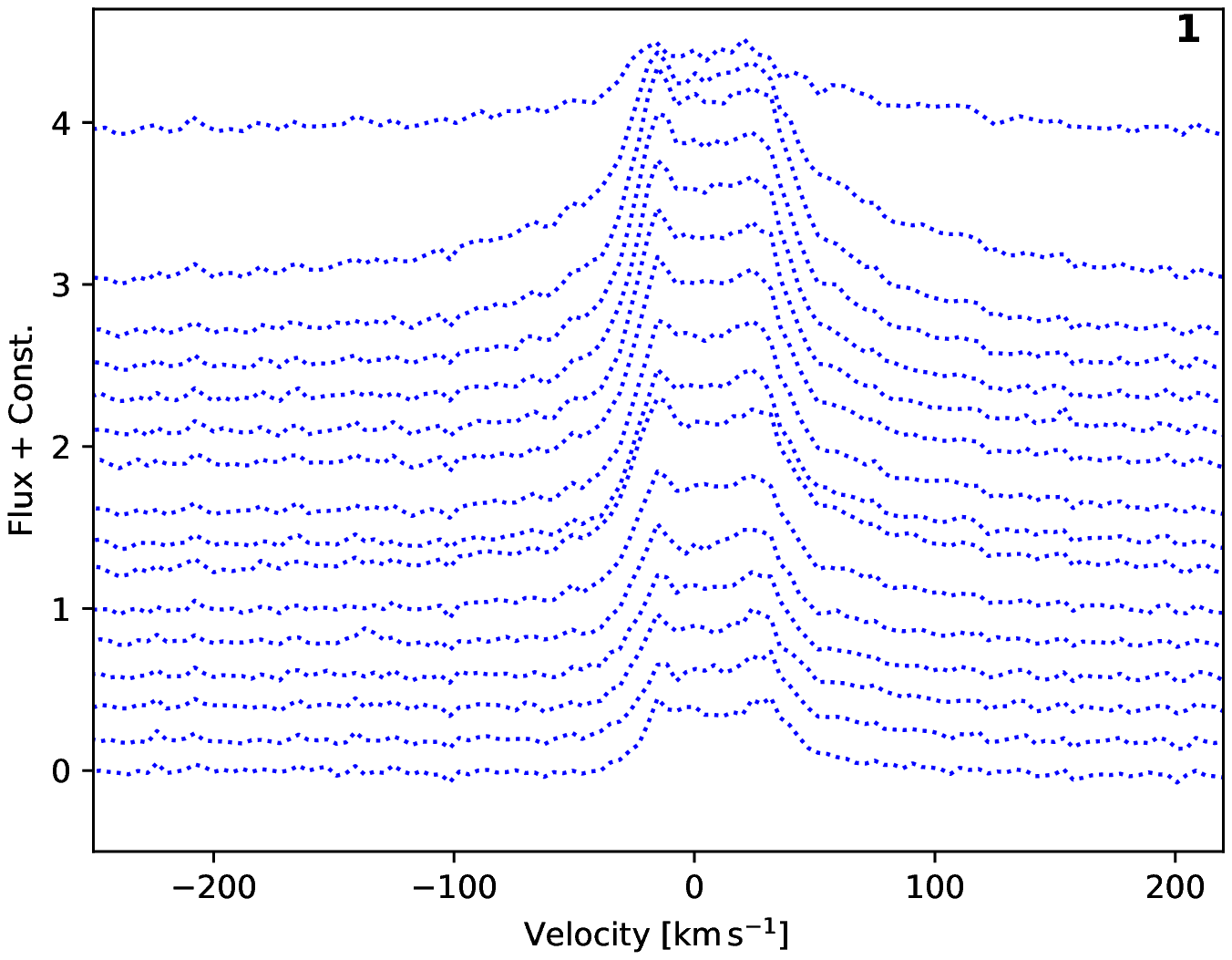}}
\vskip -0.35cm
 \subfigure
 {\includegraphics[scale=0.5]{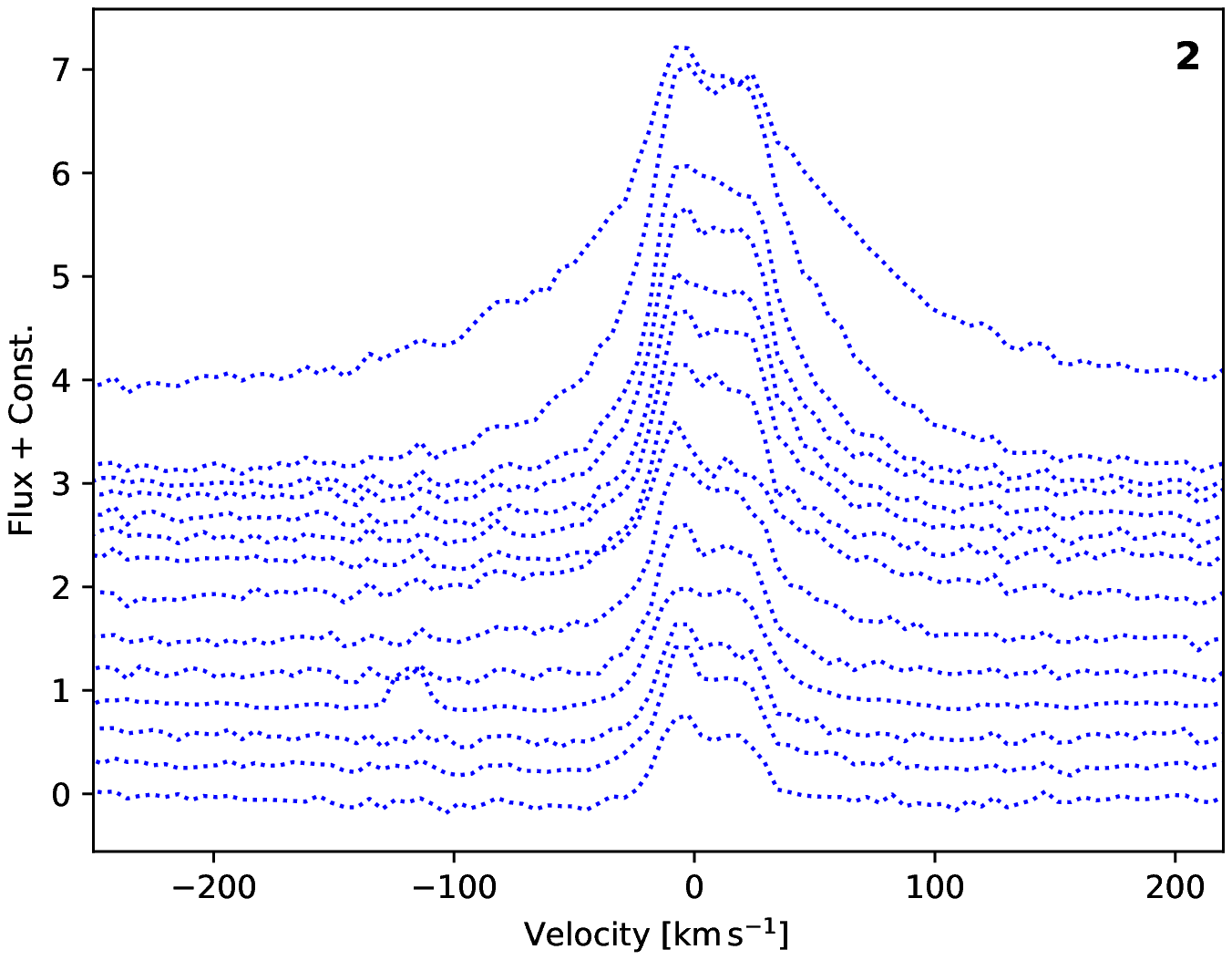}}
 \vskip -0.35cm
 \subfigure
{\includegraphics[scale=0.5]{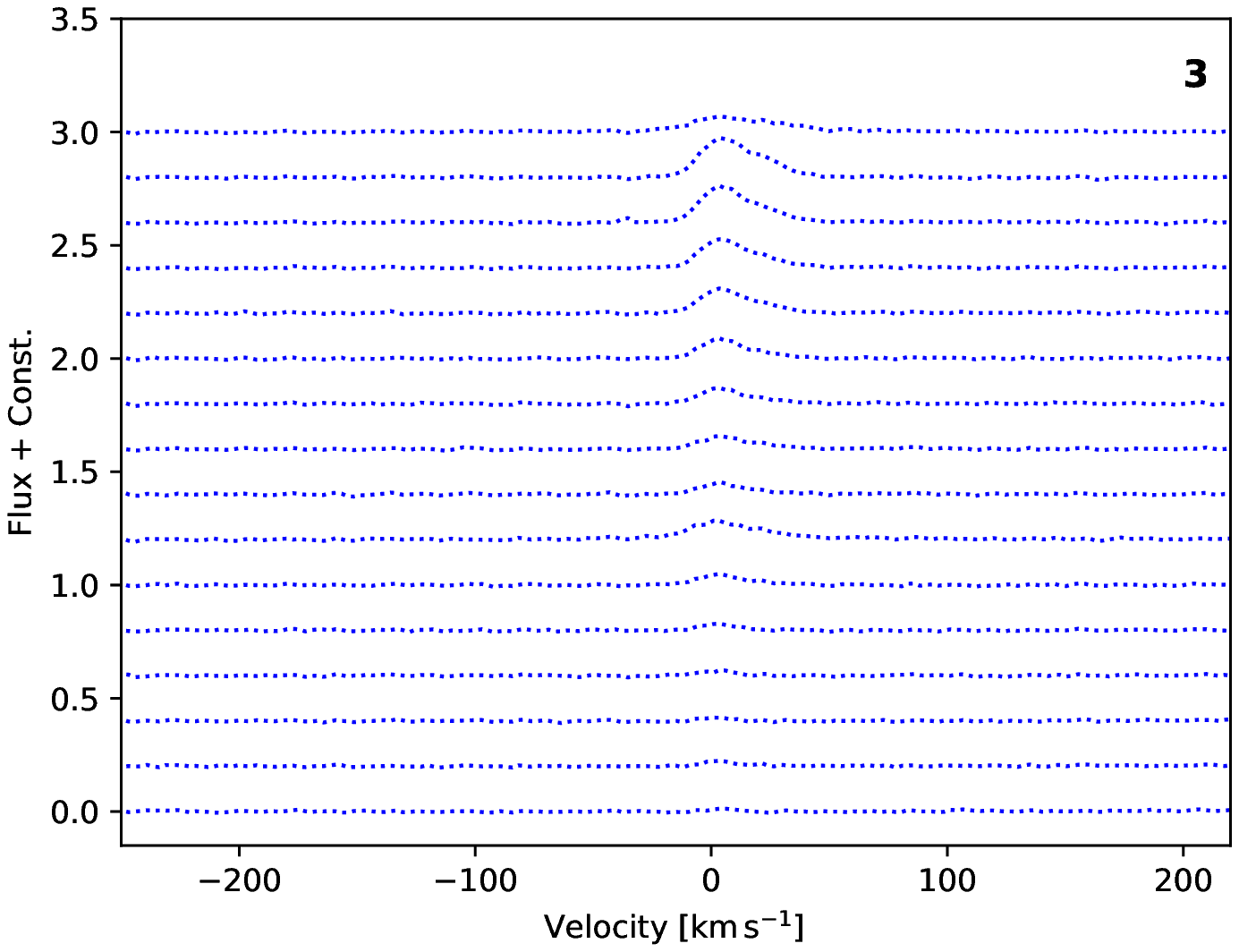}}
\caption{Evolution of the chromospheric lines: $\rm H_{\alpha}$ (panel 1), $\rm H_{\beta}$ (panel 2), and He\,I\,5876 (panel 3) during the strongest flare.}
\label{lines}

\end{figure}

 \begin{table*}[t]
\caption{Summary of the line asymmetries in $\rm H \alpha$ with the determined velocities in $\rm km\, s^{-1}$. The number of the spectra column corresponds to the number of spectra per night in which asymmetries were observed. The average velocities in the blue and red were obtained from the corresponding spectra. }
  \begin{tabular}{p{1.5cm} p{0.8cm} p{1.cm} p{1.0cm} p{1.cm} p{1.cm}cccccc}
\hline
\hline
Date &&\multicolumn{3}{c}{Velocity-Blue ($\rm km\,s^{-1}$)}&&\multicolumn{3}{c}{Velocity-Red ($\rm km\,s^{-1}$)}  && \multicolumn{2}{c}{Number of spectra}\\
&& min & max & average && min & max & average &&blue&red\\[0.5ex]
\hline
2017-05-15&&156&176&170&&157&163&161&&7&7\\
2018-02-13&&--&--&176&&--&--&--&&1&--\\
2018-02-14&&62&269&128&&100&194&154&&10&14\\
2018-03-02&&103&--&--&&135&138&137&&1&4\\
2018-03-06&&118&194&160&&141&326&219&&3&4\\
2019-02-14&&99&122&107&&--&--&--&&5&--\\
2019-02-15&&93&108&101&&105&121&113&&3&3\\
2019-02-16&&101&201&138&&192&194&193&&6&2\\
2019-02-17&&89&143&106&&114&135&127&&6&3\\
2019-02-18&&89&201&140&&102&196&150&&18&12\\
2019-02-23&&103&140&127&&--&--&--&&3&--\\
2019-02-24&&83&131&91&&114&150&132&&4&2\\
\hline
\hline
 \end{tabular}
 
 \label{tab2}
 \end{table*}
 We found 75 spectra with line asymmetries. Three large events were found, and these corresponded to nights with large flare events. However, there were nights with weak or no flares, but the spectra showed some asymmetry. This asymmetry could be interpreted as multiple chromospheric condensations that may result from unresolved low-energy flares \citep{crespo06}. To determine the velocities of the events, we subtracted the spectra for each night from the spectrum in quiescence (an average of the spectra without enhancement in $\rm H_{\alpha}$) and calculated the Doppler shift in the red and blue of the $\rm H_{\alpha}$ emission line. The maximum velocity is determined at the point where the continuum of the residual spectrum merges with the continuum of the spectrum in quiescence. In searching for CMEs, we looked for asymmetries in the blue wings of emission lines because they are often used as a possible signature. The escape velocity of \object{AD\,Leo} is $ \approx 590 \pm 11~\rm km\,s^{-1}$ , and a sign of a CME would be a blue-shifted component with a velocity greater than this value. Fig. \ref{lines} shows the evolution of the different emission line profiles during the strongest flare after subtracting the spectra in quiescence.   
There is an increase in the flux at the impulsive phase, however, the core of the lines increases faster than the wings. The emission lines also broaden during the impulsive phase of the flare and decrease with flare evolution. The broadening of the lines might be a result of the Stark broadening effect \citep{svestka72, Gizis13} and turbulent motions in the chromosphere of the star \citep{kowalski17}. In the impulsive phase,  the blue wing is more strongly enhanced, but as the flare decays, more enhancement is observed in the red wing.
We also note that the continuum does not change much in quiescence and during the flares. These flares, known as non-white flares, are common on the Sun and have previously been observed on \object{AD\,Leo}  \citep[e.g.][]{crespo06}. However, we may not completely rule out the fact that the continuum changed because there was no simultaneous photometry.
\begin{figure}
 \includegraphics[width=.5\textwidth]{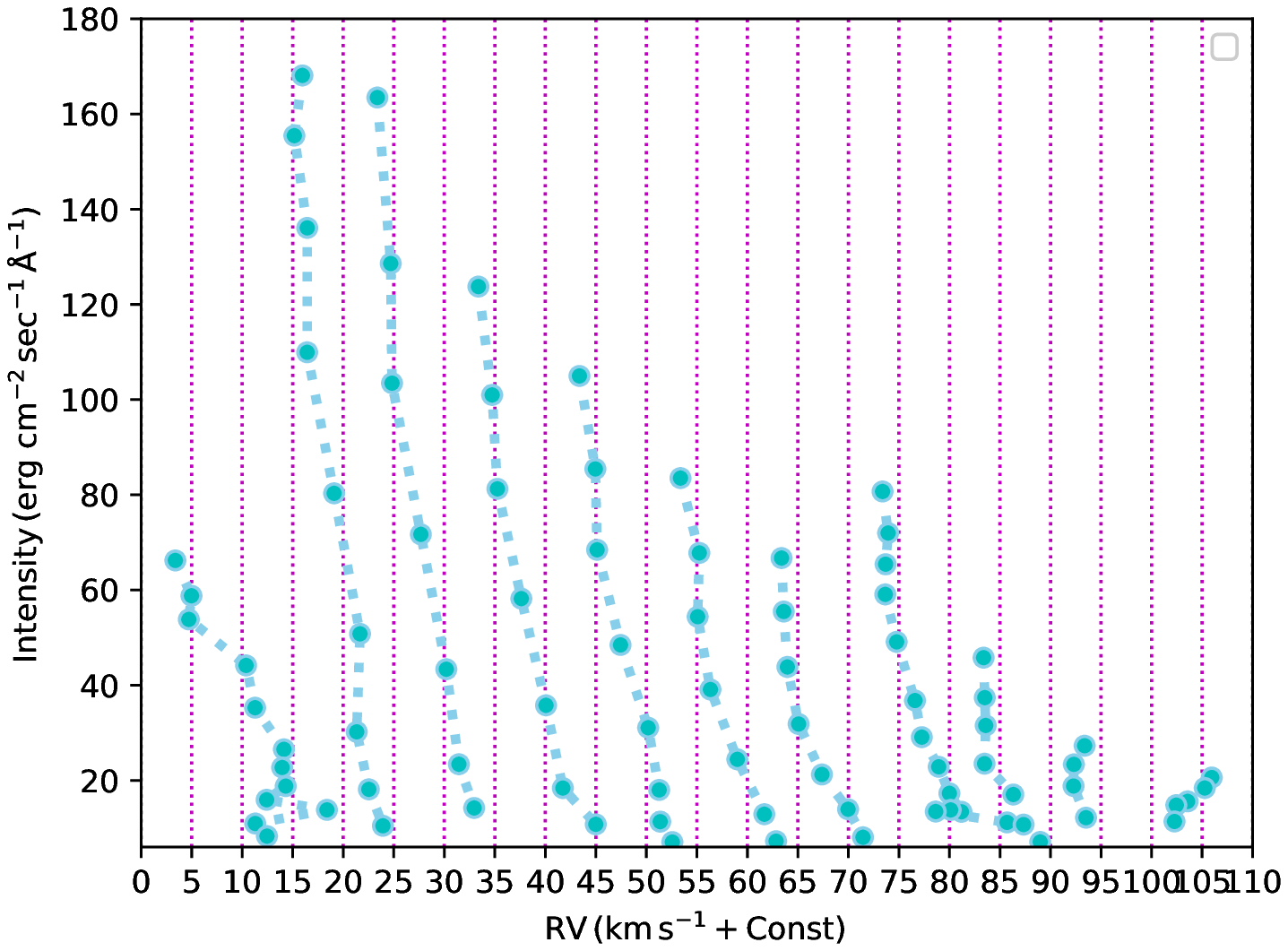}
 
\caption{\label{bisector1}Evolution of the bisector of the He\,I line during the strongest flare.}
\end{figure}
In Fig. \ref{bisector1} we show the temporal evolution of the bulk velocity (using a bisector) of the helium line during the largest flare. This line is an indicator of a very hot plasma, and the asymmetry was more pronounced in this line than in the $\rm H_{\alpha}$ and $\rm H_{\beta}$ lines. The bulk velocities of the plasma in the impulsive phase ranged between $8 - 10\, \rm km\,s^{-1}$.  

\section{Discussion}
\subsection{Line asymmetries}
In general, asymmetries can be interpreted as a signature of chromospheric plasma motions. 
Blue asymmetry is usually interpreted as upward-moving material, whereas a red asymmetry can be associated with downward motions (chromospheric condensation)   \citep{canfield90, heinzel94, crespo06, kuridze16, fuhrmeister18}. During the impulsive phase, the broad component appears blue-shifted up to $\sim 260\, \rm km\,s^{-1}$ and is then red-shifted up to $\sim 190\, \rm km\,s^{-1}$ during the gradual decay phase (see Fig.\ref{lines}).   These asymmetries have been observed in this and other M stars by several authors  \citep[e.g.~][]{houdebine90, houdebine93, gunn94, guenther97, montes99, fuhrmeister04, fuhrmeister05, crespo06, leitzinger14, vida16, fuhrmeister18}. We found that the observed velocities of the blue asymmetry events were between $80-260~ \rm km\,s^{-1}$ , which is far below the escape velocity of the star. This implies that none of these are CME events. The slow events can be attributed to three origins that we describe below.\begin{enumerate}
                                                                                                                                                                                                                                                                                                                                                                                                                                                                                                                                                                                                                                                                                                                                                                                                                                                                                                                                                                                                                                                                                                                                                                                                                                                                                                                                                                              \item[(i)] Chromospheric evaporation may be one origin, as discussed by \citet{fuhrmeister18}. During flare onset, material rises with a different velocity, producing the broad component of the line, and as the flare decays, the material flows downwards, leading to the red asymmetry that is observed  \citep{fuhrmeister08}. The authors interpreted this scenario as a chromospheric prominence that is lifted during flare onset and then rains down during  decay.  During solar flares, the velocities of chromospheric evaporation reach several tens of $\rm km\,s^{-1}$ and in extremes, even to a few hundred $\rm km\,s^{-1}$.
   \item[(ii)] The velocities we obtained are projected velocities because the propagation angle of the CMEs is unknown  \citep{leitzinger10}.                                                                                                                                                                                                                                                                                                                                                                                                                                                                                                                                                                                                                                                                                                                                                                                                                                                                                                                                                                                                                                                                                                                                                                                                                                                                                                                                                                           
     \item[(iii)] Magnetic field confinement of CMEs in very active stars is another cause \citep{drake16, alvarado18}. \citet{alvarado18} suggested that the strong magnetic fields on active stars do not allow material to leave the stellar surface. Only monster  CMEs with energies greater than $3\times 10^{32}$ erg may escape a strong field of 75\,G. Previous studies showed that \object{AD\,Leo} has a magnetic field of strength $\rm B\sim3300\,G$  \citep[][and references therein]{cranmer11} and an average magnetic field of $\sim 300-330\,\rm G$ \citep{lavail18}, which is stronger than the field (75\,G) used in the model of \citet{alvarado18}. This implies that CMEs on \object{AD\,Leo} require vast energy to escape this magnetic field. The authors further stated that eruptions following the solar flare--CME relation can only occur during flare events with energies higher than $6 \times10^{32}$ erg \citep[e.g.][]{guenther97}, but the energy in the largest event we observed was one-third of this in $\rm H_{\alpha}$. The overlying field in active stars reduces the CME speeds in comparison with the solar observations and their extrapolations. In addition to this, \citet{chen06} found in their CME velocity distribution study on the Sun that CMEs that were associated with non-active region filaments had higher eruption speeds than those with active region filaments. This shows that the magnetic field configuration plays an important role in CME propagation. 
                                                                                                                                                                                                                                                                                                                                                                                                                                                                                                                                                                                                                                                                                                                                                                                                                                                                                                                                                                                                                                                                                                                                                                                                                                                                                                                                                               \end{enumerate}
                                                                                                                                                                                                                                                                                                                                                                                                                                                                                                                                                                                                                                                                                                                                                                                                                                                                                                                                                                                                                                                                                                                                                                                                                                                                                                                                                               Red asymmetries usually occur at the impulsive phase of flares on the Sun and are driven by non--thermal downward electron beams  \citep[e.g.~][]{Ichimoto84, canfield90, heinzel94}. These red-shifted downflows in the cool lines on the Sun with velocities in the order of $ \rm \sim 100~km\,s^{-1}$ have been interpreted as solar chromospheric condensations.  
This condensation is often associated with explosive evaporation. During explosive evaporation, the chromosphere is unable to radiate the flare energy away, and this results in excess pressure that causes plasma to move downwards \citep{milligan06}. 

On the other hand, these asymmetries may not solely be attributed to chromospheric flows, but also to opacity changes occuring at different wavelengths because the velocity gradients in the chromosphere are very steep. This leads to the formation of blue and red asymmetries \citep{kuridze15}.

  The low bulk velocities obtained from the bisector of the largest event might also be an indication that material may not be rising or falling. Rather, there is a movement in the flare footpoints that causes the red and blue asymmetries. During flare reconnection, the flare footpoints mostly appear to move away from one another \citep{asai04, temmer07}. To understand this better, we need photometric observations. Thus, there is need for simultaneous photometry and spectroscopy to fully understand the processes taking place in the photosphere and the chromosphere on this star during flare events. 

 \citet{vida19} studied 21 high-resolution spectra of \object{AD\,Leo}  in their search for CMEs in late-type stars. In the 21 spectra, they found 9 events with low velocities with a maximum of  $195 ~\rm km\,s^{-1}$ in the blue and  $296 ~\rm km\,s^{-1}$ in the red, which is far below the escape velocity of \object{AD\,Leo}. Although their sample is smaller, their findings are  in agreement with our findings. This might imply that fast events are rare on this and other M stars, and solar flare--CME relations may therefore not be extended to M stars.  
 \subsection{Flares and their effect on habitability}
  Coronal mass ejections are commonly associated with flares that release energy in $10^{32}$ erg and last $\approx~ 10^{4}$ s. The total emission during the largest flare that we observed was
$2.9\times 10^{31}$ erg in $\rm H_{\beta}$ and $2.12\times 10^{32}$ erg in $\rm H\alpha$. In the case of solar flares, the energy released in soft X-rays and UV is one order of magnitude higher than that in $\rm H_{\alpha}$ \citep[see reviews by~][]{mirzoian84, haisch99, garcia00, benz10}. Because the largest flare had $2.12\times 10^{32}$ erg in $\rm H_{\alpha}$, the flux in the X-ray and UV is expected to be $ \approx 2.12\times 10^{33}$ erg.
According to \citet{somov92}, the largest solar flares have $3\times 10^{30}$ erg in $\rm H\alpha$, which means that this flare was 70 times more energetic. 
From the flare frequency distribution we obtained for flares on this star, about one event  with an energy of $3\times 10^{30}$ erg occurs per hour. On the Sun, many X class flares occur during solar maximum, but only a few or none during minimum. The average number of X class flares per year given in \citet{aschwanden12} is $\approx 6.7,$ which translates into 0.000765 events per hour. This implies that the flare activity of \object{AD\,Leo} is 1\,000 times higher than that of the Sun for very large flares.

Using the parallax of \object{AD\,Leo} from \textit{Gaia}, we obtain a distance of $4.9660\pm 0.0017$ pc, radius of 0.444$\mbox{R}_{\odot}$ , and luminosity of 0.024$\mbox{L}_{\odot}$. Following the method of determining the habitable zone boundaries in \citet{whitmire96}, we estimated its habitable zone to be at about 0.2 AU. Considering the largest flare, its intensity  would be 2\,000 times higher for a planet in the habitable zone. This means that an Earth-size planet in the habitable zone of \object{AD\,Leo} would receive more than $10^6$ times as much energy than the Earth does from solar flares. The total energy of the observed flares is $\approx 1.77\times 10^{33}$ erg during our observation time. In a year, the star would emit 35 times as much energy, which means the star would emit $\approx 6.2\times 10^{34}$ erg. When we consider a planet with the radius of the Earth in the habitable zone, the planet would receive at least $\approx 2.6\times 10^{28}$ erg per year in soft X-rays and XUV radiation.

\citet{segura10} analysed the effect of one flare with a total energy of $\sim 10^{34}$ erg that was observed on \object{AD\,Leo} in 1985. They found that the event did not necessarily present a problem for life on the surface of an Earth-like planet through the UV radiation.  However, a substantial decrement in the ozone ($\sim 94\%$) was noted, caused by protons that would be produced if there were a CME. The authors estimated that the recovery time for the effects of such a flare is about 50 years. \citet{venot16} studied the effect of the same flare on the chemical composition of two hypothetical sub-Neptune and super-Earth-like planets orbiting \object{AD\,Leo}. They found that a single flare event could irreversibly alter
the chemical composition of warm or hot planetary atmospheres, with the final steady-state being significantly different from the initial steady-state in both cases. \citet{Tilley17} modelled the effects of frequent flaring of an M dwarf on the atmosphere of an unmagnetised Earth-like planet in the habitable zone. They found that the impact of one flare (equivalent to the size of flares on \object{AD\,Leo}) per month is sufficient to drive a loss of $99.99\%$ of the ozone column within eight years, with an unlikely recovery and thus further destruction. The surface of the planet would experience a UV flux of $\sim 0.18-1\, \rm W\,m^{-2}$. The authors suggested further experiments to establish the impact of these fluxes on the onset of complex organic chemistry and life. \citet{howard18} discussed the impact of flares on the habitability of planets around Proxima Centauri. They simulated flares with energies in the range of $10^{29.5} - 10^{32.9}$ erg in the U band. They found that such events deplete the ozone by $90\%$ and the system hardly reaches a steady state with increasing time.
 This may seriously question the possibility of the existence of complex life forms around these  active stars because the planetary systems seem to experience a high loss of ozone during high-energy events, which leaves the planetary surfaces largely unprotected from UV light. Only organisms capable of tolerating extreme conditions can survive such an activity.

\section{Summary and conclusions}
 \label{sectIV}
  \object{AD\,Leo} was monitored during the period from November 2016 to February 2019 whenever it was visible for one
  week per month, and 2002 spectra were obtained and analysed. This is the largest survey of this kind on \object{AD\,Leo}. We expected to observe quite a number of flares and possibly CMEs because \citet{crespo06} obtained a flare frequency of \object{AD\,Leo}  greater than 0.71 flares per hour. However, we observed that the flare rates in 2016 and 2017 were generally low and only increased beginning in February 2018. We found 22 flares with energies of $(1.59 - 21.2)\times 10^{31}$ erg in $\rm H_{\alpha}$ and $(0.669-9.495)\times 10^{31}$ erg in $\rm H_{\beta}$. The smallest flare had an energy of $\rm 3.19\pm 0.01 \times 10^{31}\, erg$ in $\rm H_{\alpha}$. We also obtained a flare frequency of 0.092 flares per hour, which is lower than that obtained in previous studies. The low frequency might be attributed to a minimum in the stellar activity cycle as predicted by \citet{buccino14}.
On the other hand, the derived flare  frequency distribution is consistent with other previous studies, although with a steeper power-law index that is probably due to a better sampling of low-energy events.
 
 We further analysed the spectra by studying the behaviour of the chromospheric lines $\rm H_{\alpha}$, $\rm H_{\beta}$, and $\rm He\,I\,5876$ in order to search for CMEs. In this, we searched for asymmetries in the lines, which might be indicators of CMEs. We found 75 spectra with asymmetries. All these asymmetries show very low velocities in the range $60-270 ~\rm km\,s^{-1}$ , which is lower than the escape velocity of the star. These slow events and the occurrence frequency of the red asymmetry after the blue asymmetry may suggest that CMEs are rare on this star. It is therefore not appropriate to use the solar flare-CME relation to predict the occurrence of CMEs on this star. The strong overlying magnetic field on these stars seems to play an important role in suppressing their CME activity. 
 
 If CMEs are rare on stars in exoplanetary systems, they would not pose a great threat, especially regarding losses of planetary atmospheres. However, effects on exoplanetary atmospheres caused by flares might still be enormous because high-energy events occur frequently, especially at the maximum of the stellar activity cycles. The habitable zones lie close in, however. These effects are the subject of future study.

\begin{acknowledgements}

The authors acknowledge the International Science Program at Uppsala University for the financial support. The authors are also grateful to the  Th\"{u}ringer Landessternwarte Observatory, Germany for their hospitality and giving them observation time. The authors are also grateful to the anonymous referee who gave very insightful comments that have improved the quality of this work.

This work was generously supported by the Th\"{u}ringer Ministerium f\"{u}r Wirtschaft, Wissenschaft und Digitale Geselischaft. This research made use of the SIMBAD database, operated at CDS, Strasbourg, France. This research also made use of data from the European Space Agency (ESA) mission \textit{Gaia} (\url{https://www.cosmos.esa.int/gaia}), processed by the \textit{Gaia} Data Processing and Analysis Consortium (DPAC.\url{https://www.cosmos.esa.int/web/gaia/dpac/consortium}).

\end{acknowledgements}

\bibliographystyle{aa} 
\bibliography{paper}

\end{document}